# Charge Transport through Conjugated Azomethine-based Single Molecules for Optoelectronic Applications.

M. Koole,[a] R. Frisenda,[a] M. L. Petrus,[b,c] M. L. Perrin,[a] H. S.J. van der Zant,[a] T. J. Dingemans [b,*]

[a] *Kavli institute of Nanoscience, Delft University of Technology, Lorentzweg 1, 2628 CJ Delft, The Netherlands.*
[b] *Faculty of Aerospace Engineering, Delft University of Technology, Kluyverweg 1, 2926 HS Delft, The Netherlands.*
[c] *Present address: Department of Chemistry, University of Munich (LMU), Butenandtstr. 11, 81377 Munich, Germany.*
*Corresponding Author: t.j.dingemans@tudelft.nl

The single-molecule conductance of a 3-ring, conjugated azomethine was studied using the mechanically controlled breakjunction technique. Charge transport properties are found to be comparable to vinyl-based analogues; findings are supported with density functional calculations. The simple preparation and good transport properties make azomethine-based molecules an attractive class for use in polymer and single-molecule organic electronics.

**Keywords:** Azomethine, molecular electronics, mechanically controlled breakjunction, Schiff-base, organic electronics, single molecule conductance

## I. INTRODUCTION

Organic molecules have received great attention over the last two decades for their application in organic electronics such as organic light emitting diodes (OLED), organic field effect transistors (OFET), organic photovoltaics (OPV), sensors and electrochromic devices.[1-4] The synthesis of these conjugated materials generally involves transition metal catalyzed carbon-carbon coupling reactions such as Suzuki-, Stille- and Kumada-coupling.[5] However, these chemistries require stringent reaction conditions, expensive catalysts and extensive product purification.[6] The restrictive synthetic accessibility results in high costs thereby making large-scale production of these materials difficult.[7,8]

Schiff-base condensation chemistry offers a low-cost alternative route to conjugated materials. In that reaction an azomethine bond (–CH=N–, also referred to as imine) is formed on condensation of an amine with an aldehyde. This chemistry offers an attractive alternative to carbon-carbon coupling reactions since the reaction can be performed at near ambient conditions, requires no expensive catalysts, and has water as the only by-product, making purification straightforward.[9] In contrast to aliphatic azomethines,[10] conjugated azomethines are very stable against photoisomerization and also exhibit a high stability towards hydrolysis and reduction.[11,12]

Recently, conjugated azomethine-based polymers and small-molecules have shown promising performance in various organic electronics such as polymer-,[13-16] small-molecule-,[9,17] and perovskite-based[18] photovoltaics, OLEDs,[19] OFETs[20] and electrochromic devices.[21,22] In particular the low molar mass azomethines have shown good performances as a hole conducting material in bulk heterojunction devices,[23] and azomethines have been able to compete with state-of-the-art materials in a hole transporting layer in perovskite solar cells.[18] However, despite the good device performance, only very little is known about charge transport in these molecules.[20,23] Bulk measurements have shown relatively low mobilities, but it is unclear which mechanisms contribute to the low mobility.

In this work, the single-molecule conductance of an azomethine-based molecule is studied to get a better insight in its charge transport properties. The general perception is that the azomethine bond is isoelectronic to the vinyl-bond (–CH=CH–),[24] which is widely used in polymer and single-molecule organic electronics. However, the single-molecule conductance through a molecule containing azomethine bonds has not yet been studied. We performed conductance measurements of a conjugated azomethine-based single molecule using a mechanically controlled break junction (MCBJ). Our findings show that replacing the all-carbon vinyl bonds with azomethine bonds does not compromise the conductance in this class of molecules and that the azomethine-bond thus does not limit the conductance along the backbone of the molecule.

## II. EXPERIMENT

As a model compound for studying the conductance of the azomethine-bond, a symmetric azomethine, containing a thiophene and two phenyl rings was synthesized (thiophene-2,5-diylbis(N-phenylmethanimine), TYPI). At the para-position, the phenyl rings are functionalized with thiol groups, which act as the anchor to the gold electrodes (Figure 1a). To characterize the conductance of TYPI a MCBJ set-up was used.[25] TYPI was dissolved in dichloromethane at a concentration of 0.5 mMol/L.

The solution was drop cast on the MCBJ, after which the gold wire was repeatedly broken and fused at a rate of 20 cycles per minute (corresponding to an electrode displacement speed of 5 nm/s). The measurements were performed in ambient conditions at a fixed bias voltage of 100 mV.

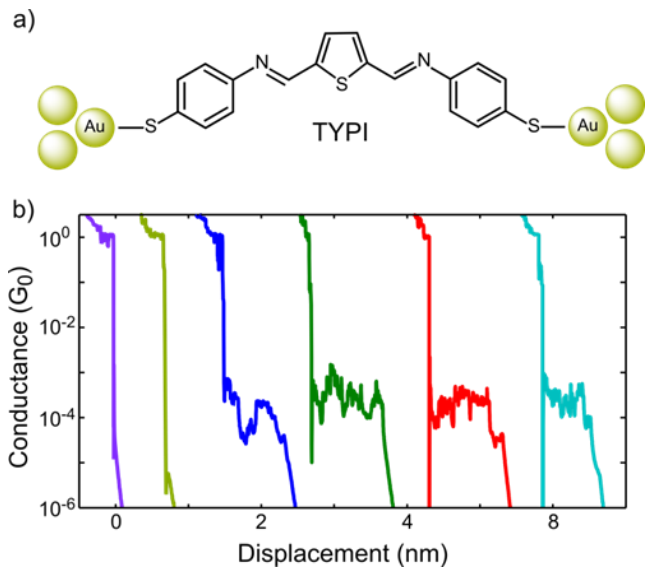

Figure 1: a) TYPI structure between gold electrodes. b) Individual breaking traces measured in the presence of TYPI. The bias voltage is 100 mV and the electrodes are withdrawn at a speed of 5 nm/s. The traces are offset along the x-axis for clarity. The conductance is plotted on a logarithmic scale.

When breaking the gold wire in the presence of TYPI we observe conductance traces as shown in Figure 1b. The conductance traces are measured as a function of electrode separation (nm). Above conductance values of 1 $G_0$ (the conductance quantum of 77 μS) the traces show decreasing steps and at approximately 1 $G_0$ a sharp drop is observed, which is caused by the rupture of the last gold few-atom connection.[26] After this point the conductance quickly drops to around $10^{-3}/10^{-4}$ $G_0$, due to the snap back of the gold apex atoms in the electrodes. Below $10^{-3}$ $G_0$ two different types of traces can be identified. The first one is an exponential decay of the conductance (linear dependence in a log-plot) to the noise level of the setup (below $10^{-6}$ $G_0$), which is characteristic of an empty junction where transport is through direct tunneling (the first two traces in Figure 1b). The second type of trace that can be identified, is the appearance of plateau-like features below $10^{-3}$ $G_0$. These plateaus persist up to 1.5 nm of stretching and show conductance fluctuations around $10^{-4}$ $G_0$. The traces with this kind of conductance-length dependence are assigned to the successful formation of a gold-molecule-gold junction.[27] After stretching for a certain distance, which is related to the length of the molecule, the conductance drops to the noise level due to breaking of the gold-molecule-gold junction.

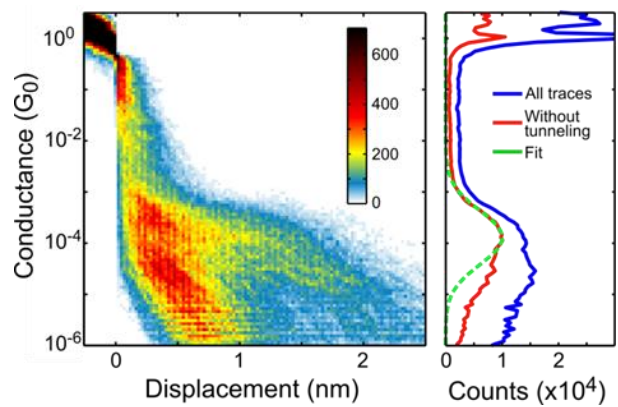

Figure 2: Conductance histograms of all breaking traces. Traces are aligned at 0.5 $G_0$. Left panel: Conductance-displacement histogram build from 2361 breaking traces like the ones shown in Figure 1b. The traces are binned logarithmically along the conductance (y) axis with 16 bins/decade and linearly binned along the displacement (x) axis with 40 bins/nm. Right panel: Conductance histogram. The blue line is built by summing the histogram in the left panel along the displacement axis. Red line is constructed excluding tunneling traces. Green dashed line is a Gaussian fit used to extract the most probable conductance value.

To determine the most probable conductance of the metal-molecule-metal junction we recorded 2361 consecutive traces, from which we build a conductance histogram. Figure 2 shows the conductance histograms built from all traces (without data selection). The zero of displacement of each trace is chosen at the point of rupture of the gold wire (first data point below 0.5 $G_0$). In the two-dimensional histogram (left side of Figure 2) a high-count region is visible above 1 $G_0$, which we assign to transport through the gold wire. Below 1 $G_0$ one can distinguish two regions of high counts. The first region drops exponentially (straight on a logarithmic scale) from $10^{-4}$ $G_0$ to the noise level extending for about 0.5-1 nm. The second region stays constant around $10^{-4}$ $G_0$ and extends up to 1.5 nm. We assign the first region to through space tunneling between the electrodes and the second region to transport through the gold-molecule-gold junctions. This second region only appears when the molecule is drop cast on the sample and is not observed in clean samples (as can be seen in the Figure 1S in the supporting info). In addition, the length of the second region approximately matches the length of a single TYPI molecule, which is 1.85 nm in DFT calculations. To extract the mean conductance of the molecular junctions we plot the 1-D conductance histogram, as shown in the right panel of Figure 2. The blue line is a histogram of the conductance including all traces, while the red line is a histogram with the tunneling region removed (further details in the supporting information). Using a Gaussian function (green dashed line) to fit the conductance distribution without tunneling, we obtain $(1.3\pm0.4)\cdot10^{-4}$ $G_0$ as most probable conduction value for TYPI.

To comment on the measured conductance value of TYPI, we compare it to the single-molecule conductance of the well-studied oligo(p-phenylene vinylene) trimer (OPV3)(Figure 3b). This molecule has approximately the same length (1.92 nm compared to 1.85 nm of TYPI) and the same basic electronic structure, which consist of three conjugated rings connected with sp2 hybridized linkers. The conductance of OPV3 with thiol anchoring groups lies between $1 \cdot 10^{-4}$ $G_0$ and $2 \cdot 10^{-4}$ $G_0$ for MCBJ setups.[28, 29] Comparing the conductance values of both TYPI and OPV3 shows that a thiophene ring with azomethine linkers has a comparable charge transport efficiency as a benzene ring with vinyl linking.

## III. THEORY

To further investigate the charge transport characteristics of TYPI, we performed density functional theory (DFT) calculations, combined with the Green's function (NEGF) formalism to obtain ground-state electronic structures and transmissions for TYPI, OPV3 and OPA3 (oligo(p-phenylene azomethine) trimer). For details concerning the calculations, see the supporting information. OPA3 is included in the calculations to differentiate between the effect that the thiophene core and azomethine bond may have on transport.

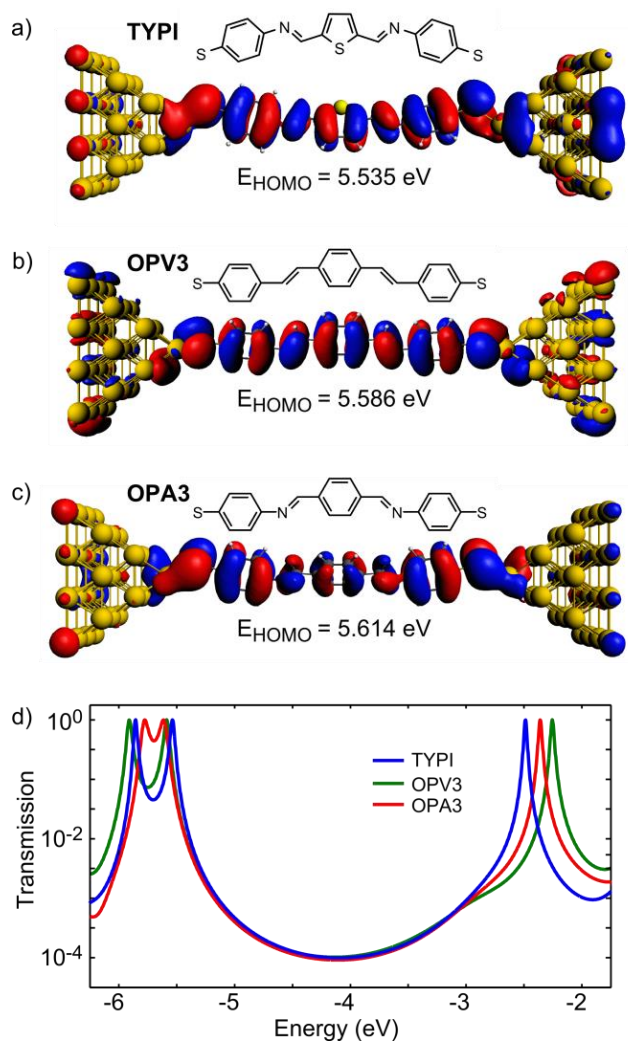

Figure 3: DFT +Σ calculations for TYPI, OPV3 and OPA3. a),b),c) Structure and hybridized HOMO orbital of respectively TYPI, OPV3 and OPA3. The energy of the HOMO is displayed underneath. d) Transmission as function of energy around the HOMO-LUMO gap for all three metal-molecule-metal junctions.

Figure 3 a-c show the isosurface of the highest occupied molecular orbital (HOMO) of TYPI, OPV3 and OPA3, respectively. It can be clearly seen that the HOMO's of all molecules have a pi character and closely resemble each other, both in shape and in energy. They extend from one sulphur atom to the other and hybridize well with the gold atoms, as can be seen by the finite density around the gold-sulphur bond and their extension throughout the entire gold electrodes. The similar electronic structure is further supported by the transmission calculations, which show similar broadening of the HOMO's indicating comparable hybridization of the molecules with the gold electrodes. Furthermore, the HOMO-LUMO gap is of comparable size, 3.1 eV for TYPI, 3.3 eV for OPV3 and 3.3 eV for OPA3.

The calculated transmission of the three molecules is shown in Figure 3d, where peaks in the transmission originate from resonant charge transport through molecular orbitals. Based on their energy, the overlapping peaks just below -5.5 eV can be identified, as the HOMO's of TYPI, OPV3 and OPA3. Focusing on the transmission in the region around the Fermi energy, which our calculations predict to be around -4.8 eV, we see that for all molecules transport is dominated by the HOMO. This is expected for thiol anchoring groups.[30, 31] Moreover, in the same energy range, the transmissions follow each other closely which is a result of the similar energy of the HOMO's and their broadening. This supports the conclusion drawn from the measurements, which is that the thiophene core with azomethine linker units has a comparable conductance to the benzene core with vinyl linking units.

The comparable conductance between OPA3 and TYPI in the calculations shows that inclusion of a thiophene core does not mask any potential negative effects the azomethine bond can have on the conductance as compared to the vinyl-linked analogue. This is further supported by comparing the transport through thiophene and phenyl units[32] which shows that despite conformational variations the conductance is approximately the same. This motivates the conclusion that the azomethine bond is of comparable conductance as a vinyl bond. The relatively poor charge mobilities that are reported for azomethines in the bulk are therefore considered to be the result of an unfavorable morphology and are not the result of a limited conductance of the azomethine bond.

## IV. CONCLUSION

In conclusion, an azomethine-based low molar mass molecule was synthesized and the conductance through single molecules was measured using a mechanically controlled break junction. The conductance was determined to be $(1.3 \pm 0.4) \cdot 10^{-4}$ $G_0$, which is comparable to reported values for OPV3, a vinyl-based analogue. This, in turn, shows that the azomethine bond exhibits good electrical conductance, making it a useful linker for the preparation of conjugated materials for electronic applications. Moreover, we believe that the simple chemistry combined with the good charge transporting properties make azomethine-based materials ideal candidates for various optoelectronic applications.

## ACKNOWLEDGEMENTS

We like to thank FOM, NWO/OCW and an ERC Advanced grant (Mols@Mols) for financial support. The work of M. L. P. was funded in part by the Dutch Polymer Institute (DPI), project #717.

## NOTES

The authors declare no competing financial interests.

## SUPPLIMENTARY INFORMATION

For further details on the chemical synthesis, experiments and DFT calculations see the supporting information.

(1) Mishra, A.; Bäuerle, P., Small Molecule Organic Semiconductors on the Move: Promises for Future Solar Energy Technology. *Angew. Chem. Int. Ed.* **2012,** 51, (9), 2020-2067.
(2) Gather, M. C.; Köhnen, A.; Meerholz, K., White Organic Light-Emitting Diodes. *Adv. Mater.* **2011,** 23, (2), 233-248.
(3) Wang, C.; Dong, H.; Hu, W.; Liu, Y.; Zhu, D., Semiconducting Π-Conjugated Systems in Field-Effect Transistors: A Material Odyssey of Organic Electronics. *Chem. Rev.* **2012,** 112, (4), 2208-2267.
(4) Thakur, V. K.; Ding, G.; Ma, J.; Lee, P. S.; Lu, X., Hybrid Materials and Polymer Electrolytes for Electrochromic Device Applications. *Adv. Mater.* **2012,** 24, (30), 4071-4096.
(5) Cheng, Y.-J.; Yang, S.-H.; Hsu, C.-S., Synthesis of Conjugated Polymers for Organic Solar Cell Applications. *Chem. Rev.* **2009,** 109, (11), 5868-5923.
(6) Nielsen, K. T.; Bechgaard, K.; Krebs, F. C., Removal of Palladium Nanoparticles from Polymer Materials †. *Macromolecules* **2005,** 38, (3), 658-659.
(7) Osedach, T. P.; Andrew, T. L.; Bulović, V., Effect of Synthetic Accessibility on the Commercial Viability of Organic Photovoltaics. *Energy Environ. Sci.* **2013,** 6, (3), 711-718.
(8) Po, R.; Bianchi, G.; Carbonera, C.; Pellegrino, A., "All That Glisters Is Not Gold": An Analysis of the Synthetic Complexity of Efficient Polymer Donors for Polymer Solar Cells. *Macromolecules* **2015,** 48, (3), 453-461.
(9) Petrus, M. L.; Bouwer, R. K. M.; Lafont, U.; Athanasopoulos, S.; Greenham, N. C.; Dingemans, T. J., Small-Molecule Azomethines: Organic Photovoltaics Via Schiff Base Condensation Chemistry. *J. Mater. Chem. A* **2014,** 2, (25), 9474-9477.
(10) Lehn, J.-M., Conjecture: Imines as Unidirectional Photodriven Molecular Motors—Motional and Constitutional Dynamic Devices. *Chem. Eur. J.* **2006,** 12, (23), 5910-5915.
(11) Bourgeaux, M.; Skene, W. G., Photophysics and Electrochemistry of Conjugated Oligothiophenes Prepared by Using Azomethine Connections. *J. Org. Chem.* **2007,** 72, (23), 8882-8892.
(12) Bourgeaux, M.; Pérez Guarin, S.; Skene, W. G., Photophysical, Crystallographic, and Electrochemical Characterization of Novel Conjugated Thiopheno Azomethines. *J. Mater. Chem.* **2007,** 17, 972-979.
(13) Hindson, J. C.; Ulgut, B.; Friend, R. H.; Greenham, N. C.; Norder, B.; Kotlewski, A.; Dingemans, T. J., All-Aromatic Liquid Crystal Triphenylamine-Based Poly(Azomethine)S as Hole Transport Materials for Opto-Electronic Applications. *J. Mater. Chem.* **2010,** 20, (5), 937-944.
(14) Iwan, A.; Palewicz, M.; Chuchmała, A.; Gorecki, L.; Sikora, A.; Mazurek, B.; Pasciak, G., Opto(Electrical) Properties of New Aromatic Polyazomethines with Fluorene Moieties in the Main Chain for Polymeric Photovoltaic Devices. *Synth. Met.* **2012,** 162, (1-2), 143-153.
(15) Iwan, A.; Boharewicz, B.; Parafiniuk, K.; Tazbir, I.; Gorecki, L.; Sikora, A.; Filapek, M.; Schab-Balcerzak, E., New Air-Stable Aromatic Polyazomethines with Triphenylamine or Phenylenevinylene Moieties Towards Photovoltaic Application. *Synth. Met.* **2014,** 195, 341-349.
(16) Petrus, M. L.; Bouwer, R. K. M.; Lafont, U.; Murthy, D. H. K.; Kist, R. J. P.; Böhm, M. L.; Olivier, Y.; Savenije, T. J.; Siebbeles, L. D. A.; Greenham, N. C.; Dingemans, T. J., Conjugated Poly(Azomethine)S Via Simple One-Step Polycondensation Chemistry: Synthesis, Thermal and Optoelectronic Properties. *Polym. Chem.* **2013,** 4, (15), 4182-4191.
(17) Moussalem, C.; Segut, O.; Gohier, F.; Allain, M.; Frère, P., Facile Access Via Green Procedures to a Material with the Benzodifuran Moiety for Organic Photovoltaics. *ACS Sustainable Chem. Eng.* **2014,** 2, (4), 1043-1048.
(18) Petrus, M. L.; Bein, T.; Dingemans, T. J.; Docampo, P., A Low Cost Azomethine-Based Hole Transporting Material for Perovskite Photovoltaics. *J. Mater. Chem. A* **2015,** 3, (23), 12159-12162.
(19) Niu, H.-J.; Huang, Y.-D.; Bai, X.-D.; Li, X., Novel Poly-Schiff Bases Containing 4,4′-Diamino-Triphenylamine as Hole Transport Material for Organic Electronic Device. *Mater. Lett.* **2004,** 58, (24), 2979-2983.
(20) Işık, D.; Santato, C.; Barik, S.; Skene, W. G., Charge-Carrier Transport in Thin Films of Π-Conjugated Thiopheno-Azomethines. *Org. Electron.* **2012,** 13, (12), 3022-3031.
(21) Mulholland, M. E.; Navarathne, D.; Petrus, M. L.; Dingemans, T. J.; Skene, W. G., Correlating on-Substrate Prepared Electrochromes with Their Solution Processed Counterparts – Towards Validating Polyazomethines as Electrochromes in Functioning Devices. *J. Mater. Chem. C* **2014,** 2, (43), 9099-9108.
(22) Ma, X.; Niu, H.; Wen, H.; Wang, S.; Lian, Y.; Jiang, X.; Wang, C.; Bai, X.; Wang, W., Synthesis, Electrochromic, Halochromic and Electro-Optical Properties of Polyazomethines with a Carbazole Core and Triarylamine Units Serving as Functional Groups. *J. Mater. Chem. C* **2015,** 3, (14), 3482-3493.
(23) Petrus, M. L.; Morgenstern, F. S. F.; Sadhanala, A.; Friend, R. H.; Greenham, N. C.; Dingemans, T. J., Device Performance of Small-Molecule Azomethine-Based Bulk Heterojunction Solar Cells. *Chem. Mater.* **2015,** 27, (8), 2990-2997.
(24) Bolduc, A.; Al Ouahabi, A.; Mallet, C.; Skene, W. G., Insight into the Isoelectronic Character of Azomethines and Vinylenes Using Representative Models: A Spectroscopic and Electrochemical Study. *J. Org. Chem.* **2013,** 78, (18), 9258-9269.
(25) Martin, C. A.; Smit, R. H. M.; van Egmond, R.; van der Zant, H. S. J.; van Ruitenbeek, J. M., A Versatile Low-Temperature Setup for the Electrical Characterization of Single-Molecule Junctions. *Rev. Sci. Instrum.* **2011,** 82, 053907.
(26) Untiedt, C.; Yanson, A. I.; Grande, R.; Rubio-Bollinger, G.; Agraït, N.; Vieira, S.; van Ruitenbeek, J. M., Calibration of the Length of a Chain of Single Gold Atoms. *Phys. Rev. B* **2002,** 66, (8), 085418.
(27) Wu, S.; González, M. T.; Huber, R.; Grunder, S.; Mayor, M.; Schönenberger, C.; Calame, M., Molecular Junctions Based on Aromatic Coupling. *Nat. Nanotechnol.* **2008,** 3, (9), 569-74.
(28) Huber, R.; González, M. T.; Wu, S.; Langer, M.; Grunder, S.; Horhoiu, V.; Mayor, M.; Bryce, M. R.; Wang, C.; Jitchati, R.; Schönenberger, C.; Calame, M., Electrical Conductance of Conjugated Oligomers at the Single Molecule Level. *J. Am. Chem. Soc.* **2008,** 130, (3), 1080-1084.
(29) Arroyo, C. R.; Frisenda, R.; Moth-Poulsen, K.; Seldenthuis, J. S.; Bjørnholm, T.; van der Zant, H. S., Quantum Interference Effects at Room Temperature in Opv-Based Single-Molecule Junctions. *Nanoscale Res. Lett.* **2013,** 8, (1), 234.
(30) Kim, B.; Beebe, J. M.; Jun, Y.; Zhu, X. Y.; Frisbie, C. D., Correlation between Homo Alignment and Contact Resistance in Molecular Junctions: Aromatic Thiols Versus Aromatic Isocyanides. *J. Am. Chem. Soc.* **2006,** 128, (15), 4970-4971.
(31) Cohen, R.; Stokbro, K.; Martin, J. M. L.; Ratner, M. A., Charge Transport in Conjugated Aromatic Molecular Junctions: Molecular Conjugation and Molecule-Electrode Coupling. *J. Phys. Chem. C* **2007,** 111, (40), 14893-14902.
(32) Dell, E. J.; Capozzi, B.; DuBay, K. H.; Berkelbach, T. C.; Moreno, J. R.; Reichman, D. R.; Venkataraman, L.; Campos, L. M., Impact of Molecular Symmetry on Single-Molecule Conductance. *J. Am. Chem. Soc.* **2013,** 135, (32), 11724-11727.